\documentclass[prd,twocolumn,groupedaddress,showpacs,nofootinbib]{revtex4}
\usepackage{graphicx}
\usepackage{dcolumn}
\usepackage{amssymb}
\usepackage{mathrsfs}
\usepackage{amsmath}
\usepackage{epsfig}
\usepackage[dvips]{color}
\usepackage{hhline}

\begin{document}

\title{Significant neutrinoless double beta decay with quasi-Dirac neutrinos}

\author{Pei-Hong Gu}
\email{peihong.gu@mpi-hd.mpg.de}

\affiliation{Max-Planck-Institut f\"{u}r Kernphysik, Saupfercheckweg
1, 69117 Heidelberg, Germany}

\begin{abstract}

A significant signal of neutrinoless double beta decay can be
consistent with the existence of light quasi-Dirac neutrinos. To
demonstrate this possibility, we consider a realistic model where
the neutrino masses and the neutrinoless double beta decay can be
simultaneously generated after a Peccei-Quinn symmetry breaking.

\end{abstract}


\maketitle

\section{Introduction}

The process of neutrinoless double beta decay \cite{furry1939} is
motivated by the conjecture of Majorana neutrinos
\cite{majorana1937} because they both need a lepton number violation
of two units. Specifically, the amplitude of the neutrinoless double
beta decay is proportional to the Majorana mass of the electron
neutrino. For a Majorana neutrino, its small mass \cite{amster2008}
can be naturally understood by the seesaw mechanism
\cite{minkowski1977,yanagida1979,grs1979,glashow1980,ms1980,sv1980,mw1980,cl1980,lsw1981,ms1981}.
There have been a lot of works (e.g.
\cite{dknot1981,zk1981,zk1981-2,hkk1996,pp2008,rodejohann2010})
studying the standard scenario of the neutrinoless double beta
decay. The electron neutrino can mediate a neutrinoless double beta
decay even if it is a quasi-Dirac neutrino \cite{valle1983,vs1983}.
However, the quasi-Dirac neutrino can not induce a significant
neutrinoless double beta decay since its Majorana mass is much
smaller than the tiny neutrino mass itself. Alternatively, the
neutrinoless double beta decay may not be directly governed by the
Majorana neutrino masses
\cite{ms1981,sv1982,mohapatra1986,hkk1995,prv2003,akp2009,bj2010,tnnsv2011}.
In this case, the lepton number violation for the neutrinoless
double beta decay eventually will result in a Majorana neutrino mass
term at loop level according to the Schechter-Valle theorem
\cite{sv1982,nieves1984,takasugi1984}.

In this paper, we shall consider an interesting scenario where the
quasi-Dirac neutrinos and the neutrinoless double beta decay are
simultaneously generated after a Peccei-Quinn \cite{pq1977} (PQ)
symmetry breaking. In our model, the PQ symmetry breaking will lead
to a lepton number violating interaction between two TeV-scale
colored scalars. This lepton number violation can induce a
neutrinoless double beta decay through the Yukawa couplings of the
colored scalars to the right-handed up-type quarks, down-type quarks
and charged leptons. The Majorana masses of the left-handed
neutrinos can be produced at four-loop order. However, their
magnitude is highly suppressed by the loop and chirality suppression
factors. The loop-induced Majorana masses hence are too small to
explain the neutrino oscillations \cite{amster2008}. Fortunately,
the Dirac masses between the left- and right-handed neutrinos can
arrive at the desired values as our model accommodates a variant
seesaw mechanism \cite{gh2006}.

\section{The model}

We extend the standard model (SM) gauge symmetry by a PQ global
symmetry. The field content is summarized in Table \ref{field},
where the gauge-singlet fermions $\nu_{R_i^{}}^{}$ (right-handed
neutrinos), the isodoublet scalar $\phi_\nu^{}$, the isosinglet
scalars $\delta$, $\omega$ and the gauge-singlet scalar $\chi$,
respectively, carry the baryon and lepton numbers $(0,2)$, $(0,1)$,
$(\frac{1}{3},1)$, $(-\frac{2}{3},0)$ and $(0,-2)$, the
gauge-singlet scalar $\sigma$ and the isodoublet scalars
$\phi_u^{}$, $\phi_d^{}$ have no baryon number or lepton number,
while the SM quarks $q_{L_i^{}}^{}$, $u_{R_i^{}}^{}$,
$d_{R_i^{}}^{}$ and leptons $l_{L_i^{}}^{}$, $e_{R_i^{}}^{}$ carry
the usual baryon number $\frac{1}{3}$ or lepton number $1$. The
global symmetry of baryon number is exactly conserved while the
global symmetry of lepton number is softly broken. Under the imposed
gauge and global symmetries, the allowed Yukawa interactions only
include
\begin{eqnarray}
\label{yukawa}
\mathcal{L}_Y^{}&=&-y_{u_{ij}^{}}^{}\bar{q}_{L_i^{}}^{}\tilde{\phi}_u^{}u_{R_j^{}}^{}
-y_{d_{ij}^{}}^{}\bar{q}_{L_i^{}}^{}\phi_d^{}d_{R_j^{}}^{}
-y_{e_{ij}^{}}^{}\bar{l}_{L_i^{}}^{}\phi_d^{}e_{R_j^{}}^{}\nonumber\\
&&
-y_{\nu_{ij}^{}}^{}\bar{l}_{L_i^{}}^{}\tilde{\phi}_\nu^{}\nu_{R_j^{}}^{}
-h_{ij}^{}\delta \bar{u}_{R_i^{}}^{}e_{R_j^{}}^c-f_{ij}^{}\omega
\bar{d}_{R_i^{}}^c d_{R_j^{}}^{}\nonumber\\
&&+\textrm{H.c.}\,.
\end{eqnarray}
As for the other gauge invariant Yukawa and mass terms,
\begin{eqnarray}
\mathcal{L}'^{}&=&-h'^{}_{ij}\delta \bar{q}_{L_i^{}}^c
i\tau_2^{}q_{L_j^{}}^{}-h''^{}_{ij}\delta \bar{u}_{R_i^{}}^c
d_{R_j^{}}^{}-h'''^{}_{ij}\delta \bar{q}_{L_i^{}}^{}
i\tau_2^{}l_{L_j^{}}^c\nonumber\\
&&-h''''^{}_{ij}\delta
\bar{d}_{R_i^{}}^{}\nu_{R_j^{}}^c-\frac{1}{2}g_{ij}^{}\chi\bar{\nu}_{R_i^{}}^{}\nu_{R_j^{}}^c
-\frac{1}{2}g'^{}_{ij}\chi\bar{\nu}_{R_i^{}}^c\nu_{R_j^{}}^{}\nonumber\\
&&-\frac{1}{2}M_{ij}^{}\bar{\nu}_{R_i^{}}^{}\nu_{R_j^{}}^c+\textrm{H.c.}\,,
\end{eqnarray}
they are forbidden by the conservation of the baryon number, the
lepton number or the PQ charge. We will not write down the full
scalar potential. Instead, we only give the following terms,
\begin{eqnarray}
\label{potential} V \supset \rho \sigma\phi^\dagger_d\phi_u^{}
+\xi\chi\sigma^2_{}+\eta\chi\phi^\dagger_\nu\phi_u^{}+\kappa
\chi\omega \delta^2_{} +\textrm{H.c.}\,,
\end{eqnarray}
which are essential to our demonstration. Clearly, the $\xi$-term
and $\eta$-term softly break the lepton number. Note that there are
no other lepton number violating terms. For convenience and without
loss of generality, we will take the parameters $\rho$, $\xi$,
$\eta$ and $\kappa$ to be real by a proper phase rotation.

\begin{table}
\begin{center}
\begin{tabular}{|c|c|c|}  \hline &&\\[0.5mm]
~&~$SU(3)_{\textrm{c}}^{}\times SU(2)_{\textrm{L}}^{}\times
U(1)_{\textrm{Y}}^{}$~
& ~$U(1)_{\textrm{PQ}}^{}$~ \\
&&\\[-1.0mm]
\hline &&\\[-0.3mm]~~$q_{L_i^{}}^{}=\left(\begin{array}{c}u_{L_i^{}}^{}\\
[2mm]d_{L_i^{}}^{}\end{array}\right)$~~ &
$(\textbf{3},~~\textbf{2},\,+\frac{1}{6})$ &$~~0$\\
&& \\[-2.2mm]
\hline  &&\\[-0.3mm]$d_{R_i^{}}^{}$ &
$(\textbf{3},~~\textbf{1},\,-\frac{1}{3})$&$+1$\\
&& \\[-2.2mm]
\hline  &&\\ [-0.3mm] $u_{R_i^{}}^{}$ &
$(\textbf{3},~~\textbf{1},\,+\frac{2}{3})$ & $~~0$\\
&& \\[-2.2mm]
\hline  && \\[-0.3mm]$l_{L_i^{}}^{}=\left(\begin{array}{c}\nu_{L_i^{}}^{}\\
[2mm]e_{L_i^{}}^{}\end{array}\right)$ &
$(\textbf{1},~~\textbf{2},\,-\frac{1}{2})$ & $-1$\\
&& \\[-2.2mm]
\hline  &&\\[-0.3mm]$ e_{R_i^{}}^{}$ &
$(\textbf{1},~~\textbf{1},\,-1)$ & $~~0$\\
&& \\[-2.2mm]
\hline  &&\\[-0.3mm]$ \nu_{R_i^{}}^{}$ &
$(\textbf{1},~~\textbf{1},\,~~0)$ & $+1$\\
&& \\[-2.2mm]
\hline  &&\\[-0.3mm]$ \delta$ &
$(\textbf{3},~~\textbf{1},\,-\frac{1}{3})$ & $~~0$\\
&& \\[-2.2mm]
\hline  &&\\[-0.3mm]$ \omega$ &
$(\textbf{6},~~\textbf{1},\,+\frac{2}{3})$ & $-2$\\
&&\\[-2.2mm]
\hline  &&\\[-0.3mm]$ \phi_u^{}=\left(\begin{array}{c}\phi_u^+\\
[2mm]\phi_u^0\end{array}\right)$ &
$(\textbf{1},~~\textbf{2},\,+\frac{1}{2})$ & $~~0$\\
&& \\[-2.2mm]
\hline  &&\\[-0.3mm]$ \phi_d^{}=\left(\begin{array}{c}\phi_d^+\\
[2mm]\phi_d^0\end{array}\right)$ &
$(\textbf{1},~~\textbf{2},\,+\frac{1}{2})$ & $-1$\\
&& \\[-2.2mm]
\hline  &&\\[-0.3mm]$\phi_\nu^{}=\left(\begin{array}{c}\phi_\nu^+\\
[2mm]\phi_\nu^0\end{array}\right)$ &
$(\textbf{1},~~\textbf{2},\,+\frac{1}{2})$ &$+2$\\
&& \\[-2.2mm]
\hline  &&\\[-0.3mm]$\sigma$ &
$(\textbf{1},~~\textbf{1},\,~~0)$ &$-1$ \\
&& \\[-2.2mm]
\hline  &&\\[-0.3mm]$\chi$ &
$(\textbf{1},~~\textbf{1},\,~~0)$ &$+2$ \\
[2.0mm]\hline
\end{tabular}
\caption{\label{field} Field content. }
\end{center}
\end{table}

The scalars $\sigma$ and $\phi_u^{}$ are responsible for the PQ
symmetry breaking and the electroweak symmetry breaking,
respectively. The vacuum expectation values (VEVs)
$\langle\sigma\rangle$ and $\langle\phi_u^0\rangle$ can induce the
VEVs of the scalars $\phi_d^{}$, $\phi_\nu^{}$ and $\chi$,
\begin{eqnarray}
\label{vevd} \langle\phi_d^0\rangle&\simeq&
-\frac{\rho\langle\sigma\rangle\langle\phi_u^0\rangle}{m_{\phi_d^0}^2}\,,\\
[3mm]
\label{vevlv}\langle\chi\rangle&\simeq&-\frac{\xi\langle\sigma\rangle^2_{}}{m_\chi^2}\,,\\
[3mm] \label{vevnu} \langle\phi_\nu^0\rangle&\simeq&
-\frac{\eta\langle\chi\rangle\langle\phi_u^0\rangle}{m_{\phi_\nu^0}^2}\,,
\end{eqnarray}
like the type-II seesaw \cite{sv1980,mw1980,cl1980,lsw1981,ms1981}.
Clearly, our model can result in an invisible
\cite{kim1979,svz1980,dfs1981,zhitnitsky1980} axion
\cite{pq1977,weinberg1978,wilczek1978} to solve the strong CP
problem. We should keep in mind the constraints on the VEVs
\cite{amster2008},
\begin{eqnarray}
&10^9_{}\,\textrm{GeV}\lesssim\langle\sigma\rangle\lesssim10^{12}_{}\,\textrm{GeV}\,,&\\
[3mm]
&\sqrt{\langle\phi_u^0\rangle^2_{}+\langle\phi_d^0\rangle^2_{}+\langle\phi_\nu^0\rangle^2_{}}\simeq
174\,\textrm{GeV}\,.&
\end{eqnarray}

\section{Quasi-Dirac neutrinos}

When the isodoublet scalars $\phi_u^{}$, $\phi_d^{}$ and
$\phi_\nu^{}$ acquire their VEVs, the up-type quarks, the down-type
quarks, the charged leptons and the neutrinos can obtain their Dirac
masses,
\begin{eqnarray}
\mathcal{L}&\supset&
-m_{u_{ij}^{}}^{}\bar{u}_{L_i^{}}^{}u_{R_j^{}}^{}
-m_{d_{ij}^{}}^{}\bar{d}_{L_i^{}}^{}d_{R_j^{}}^{}
-m_{e_{ij}^{}}^{}\bar{e}_{L_i^{}}^{}e_{R_j^{}}^{}\nonumber\\
&&
-m_{\nu_{ij}^{}}^{}\bar{\nu}_{L_i^{}}^{}\nu_{R_j^{}}^{}+\textrm{H.c.}
\end{eqnarray}
with
\begin{eqnarray}
m_{u_{ij}^{}}^{}&=&y_{u_{ij}^{}}^{}\langle\phi_u^0\rangle\,,\\
m_{d_{ij}^{}}^{}&=&y_{d_{ij}^{}}^{}\langle\phi_d^0\rangle\,,\\
m_{e_{ij}^{}}^{}&=&y_{e_{ij}^{}}^{}\langle\phi_d^0\rangle\,,\\
\label{numass}
m_{\nu_{ij}^{}}^{}&=&y_{\nu_{ij}^{}}^{}\langle\phi_\nu^0\rangle\,.
\end{eqnarray}
To generate the known charged fermion masses \cite{amster2008}, we
should perform
\begin{eqnarray}
\langle\phi_u^0\rangle&=&\mathcal{O}(100\,\textrm{GeV})\,,\\
\langle\phi_d^0\rangle&=&\mathcal{O}(\textrm{1--10\,GeV})
\left[\frac{\langle\phi_u^0\rangle}{\mathcal{O}(100\,\textrm{GeV})}\right]
\left[\frac{\rho}{\mathcal{O}(0.1\,m_{\phi_d^0}^{})}\right]\nonumber\\
&&\times\left[\frac{m_{\phi_d^0}^{}}{\mathcal{O}(\langle\sigma\rangle)}\right]^{-2}_{}
\left[\frac{\langle\sigma\rangle}{\mathcal{O}(10^{12}_{}\,\textrm{GeV})}\right]\,.
\end{eqnarray}

\begin{figure*}
\vspace{7.9cm} \epsfig{file=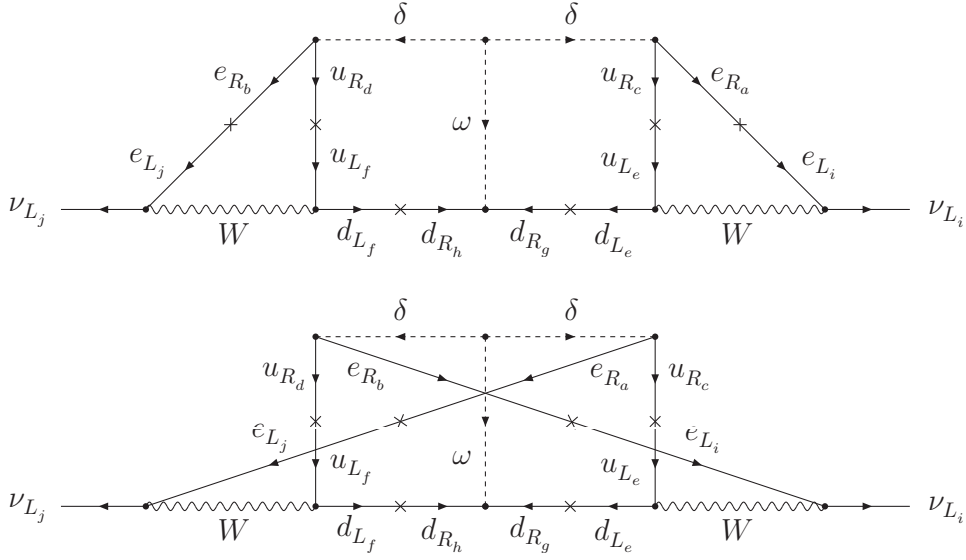, bbllx=3.8cm, bblly=6.0cm,
bburx=13.8cm, bbury=16cm, width=8cm, height=8cm, angle=0, clip=0}
\vspace{-8.2cm} \caption{\label{majorana} The left-handed neutrinos
obtain their Majorana masses at four-loop order.}
\end{figure*}

In the presence of the $\kappa$-term in the potential
(\ref{potential}), the VEV $\langle\chi\rangle$ given by Eq.
(\ref{vevlv}) will result in a lepton number violating interaction
between the colored scalars $\delta$ and $\omega$,
\begin{eqnarray}
\label{lnviolation} \mathcal{L}\supset -\mu\omega \delta^2_{}
+\textrm{H.c.}\quad\textrm{with}\quad \mu=\kappa
\langle\chi\rangle\,.
\end{eqnarray}
Remarkably, the VEV $\langle\chi\rangle$ can be close to the TeV
scale for a nice parameter choice,
\begin{eqnarray}
\langle\chi\rangle&\simeq&\mathcal{O}(10^5_{}\,\textrm{GeV})
\left[\frac{\langle\sigma\rangle}{\mathcal{O}(10^{12}_{}\,\textrm{GeV})}\right]^2_{}
\left[\frac{\xi}{\mathcal{O}(0.1\,m_\chi^{})}\right]
\nonumber\\&&\times
\left[\frac{m_{\chi}^{}}{\mathcal{O}(m_{\textrm{Pl}}^{})}\right]^{-2}_{}\,.
\end{eqnarray}
Here $m_{\textrm{Pl}}^{}\simeq 2.43\times 10^{18}_{}\,\textrm{GeV}$
is the reduced Planck mass. We then can immediately read
\begin{eqnarray}
\mu=\mathcal{O}(10\,\textrm{TeV})\quad\textrm{for}\quad\kappa=\mathcal{O}(0.1)\,.
\end{eqnarray}
Due to the lepton number violation (\ref{lnviolation}), the
left-handed neutrinos can obtain their Majorana masses through the
four-loop diagrams as shown in Fig. \ref{majorana}. We can rapidly
estimate the loop-induced Majorana masses as below,
\begin{eqnarray}
\mathcal{L}\supset -\frac{1}{2} \delta
m_{\nu_{ij}^{}}^{}\bar{\nu}_{L_i^{}}^{}\nu_{L_j^{}}^c+\textrm{H.c.}\,,
\end{eqnarray}
where
\begin{eqnarray}
\delta m_{\nu_{ij}^{}}^{}&\simeq&
\frac{4}{(16\pi^2)^4_{}}\left(\frac{g}{\sqrt{2}}\right)^4_{}\frac{\mu}{m_\omega^2
m_\delta^4}\sum_{abcdefgh}f_{gh}^{}h_{ca}^{}h_{db}^{}\nonumber\\
&&\times (m_{e_{ia}^{}}^{} m_{e_{jb}^{}}^{}+m_{e_{ib}^{}}^{}
m_{e_{ja}^{}}^{})m_{u_{ce}^{}}^{}m_{u_{df}^{}}^{} m_{d_{ge}^{}}^\ast
m_{d_{hf}^{}}^\ast\nonumber\\
&&
\end{eqnarray}
with $g$ being the $SU(2)_{\textrm{L}}^{}$ gauge coupling. It is
easy to find that the loop-induced Majorana masses should have an
upper bound,
\begin{eqnarray}
\delta m_{\nu_{ij}^{}}^{}\lesssim
\frac{g^4_{}}{2^{15}_{}\pi^8}\frac{\mu m_\tau^2 m_t^2 m_b^2
}{m_\omega^2 m_\delta^4} \,,
\end{eqnarray}
which yields
\begin{eqnarray}
\delta m_{\nu_{ij}^{}}^{}&\lesssim&
\mathcal{O}(10^{-8}_{}\,\textrm{eV})\left[\frac{\mu}{\mathcal{O}(10\,\textrm{TeV})}\right]
\left[\frac{m_\omega^{}}{\mathcal{O}(\textrm{TeV})}\right]^{-2}_{}\nonumber\\
&&\times
\left[\frac{m_\delta^{}}{\mathcal{O}(\textrm{TeV})}\right]^{-4}_{}\,.
\end{eqnarray}
In the above numerical estimation we have taken into account
$g=0.653$, $m_\tau^{}=1.78\,\textrm{GeV}$,
$m_t^{}=171.2\,\textrm{GeV}$ and $m_b^{}=4.20\,\textrm{GeV}$
\cite{amster2008}. Therefore, the Majorana masses of the left-handed
neutrinos are far below the values to explain the phenomenon of
neutrino oscillations \cite{amster2008}.

\begin{figure}
\vspace{8.5cm} \epsfig{file=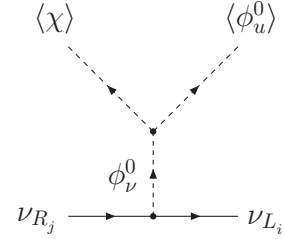, bbllx=3.0cm, bblly=6.0cm,
bburx=13.0cm, bbury=16cm, width=8cm, height=8cm, angle=0, clip=0}
\vspace{-12.5cm} \caption{\label{dirac} The left- and right-handed
neutrinos obtain their Dirac masses at tree level.}
\end{figure}

We then check the Dirac masses between the left- and right-handed
neutrinos. It is striking that the Dirac neutrino masses
(\ref{numass}) can arrive at the desired level without fine tuning
the Yukawa couplings $y_{\nu_{ij}^{}}^{}$ \cite{gh2006} because the
VEV $\langle\phi_\nu^0\rangle$ given by Eq. (\ref{vevnu}) is just
\begin{eqnarray}
\langle\phi_\nu^0\rangle&\simeq&\mathcal{O}(0.01-0.1\,\textrm{eV})
\left[\frac{\langle\phi_u^{}\rangle}{\mathcal{O}(100\,\textrm{GeV})}\right]
\left[\frac{\langle\chi\rangle}{\mathcal{O}(10^5_{}\,\textrm{GeV})}\right]
\nonumber\\
&&\times \left[\frac{\eta}{\mathcal{O}(m_{\phi_\nu^0}^{})}\right]
\left[\frac{m_{\phi_\nu^0}^{}}{\mathcal{O}(m_{\textrm{Pl}}^{})}\right]^{-2}_{}\,.
\end{eqnarray}
Obviously, the generation of Dirac neutrino masses retains the
essence of the conventional type-II seesaw
\cite{sv1980,mw1980,cl1980,lsw1981,ms1981}, as shown in Fig.
\ref{dirac}.

We now can conclude that the neutrinos in our model are quasi-Dirac
particles.

\section{Neutrinoless double beta decay}

\begin{figure}
\vspace{4.9cm} \epsfig{file=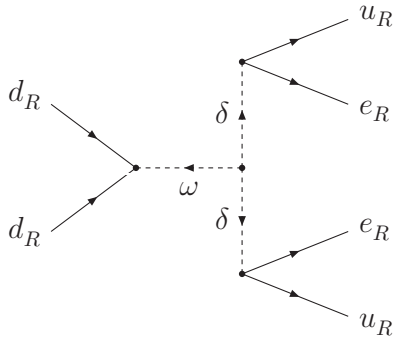, bbllx=4.6cm, bblly=6.0cm,
bburx=14.6cm, bbury=16cm, width=8cm, height=8cm, angle=0, clip=0}
\vspace{-7.7cm} \caption{\label{neutrinoless} The neutrinoless
double beta decay is mediated by the colored scalars. }
\end{figure}

We can integrate out the heavy colored scalars $\delta$ and $\omega$
from Eqs. (\ref{yukawa}) and (\ref{lnviolation}). The leading
dimension-9 operator should be
\begin{eqnarray}
\mathcal{O}_9^{}&=&\frac{\mu}{m_\omega^2 m_\delta^4}
f_{ij}^{}h_{kl}^{}h_{mn}\bar{d}_{R_i^{}}^c
d_{R_j^{}}^{}\bar{u}_{R_k^{}}^{}e_{R_l^{}}^c\bar{u}_{R_m^{}}^{}e_{R_n^{}}^c+\textrm{H.c.}\,,\nonumber\\
&&
\end{eqnarray}
which can result in a neutrinoless double beta decay as shown in
Fig. \ref{neutrinoless}. With the cubic coupling
$\mu=\mathcal{O}(10\,\textrm{TeV})$ and the Yukawa couplings
$h_{11}^{},\,f_{11}^{}\lesssim\mathcal{O}(1)$, the induced
neutrinoless double beta decay can be verified by the ongoing and
planned experiments \cite{barabash2011} if the colored scalars are
at the TeV scale, i.e.
$m_\delta^{},\,m_\omega^{}=\mathcal{O}(\textrm{TeV})$. Note that we
have the flexibility to avoid the stringent constraints from other
rare processes \cite{amster2008} by choosing the values of the
Yukawa couplings $f_{ij}^{},\,h_{ij}^{}~[(i,j)\neq (1,1)]$. With the
TeV-scale colored scalars, we can also expect to check our scenario
at colliders such as the LHC.

\section{Conclusion}

In this work, we have demonstrated a scenario where the neutrinoless
double beta decay can be observed by the forthcoming experiments
within the next few years even if the light neutrinos are
quasi-Dirac particles. In our model, a spontaneous PQ symmetry
breaking will induce a lepton number violating interaction between
two colored scalars. Because the colored scalars do not have any
Yukawa interactions involving the left-handed fermions or the
right-handed neutrinos, the lepton number violation can
significantly result in the neutrinoless double beta decay at tree
level while can negligibly contribute to the left-handed Majorana
neutrino mass at four-loop order. Through a variant seesaw
mechanism, the quasi-Dirac neutrinos can naturally obtain the
desired masses to explain the neutrino oscillations.

\vspace{3mm}

\textbf{Acknowledgement}: I thank Manfred Lindner for hospitality at
Max-Planck-Institut f\"{u}r Kernphysik. This work is supported by
the Alexander von Humboldt Foundation.

\end{document}